\documentstyle[12pt,aasms4,psfig]{article} 
\footnotesep \baselineskip      
\def\lsim{\hbox{\raise.35ex\rlap{$<$}\lower.6ex\hbox{$\sim$}\ }}
\def\ein{{\it Einstein}}

\def\rosat{{\it ROSAT}}

\def\asca{{\it ASCA}}
\def\egret{{\it EGRET}}
\def\myarcmin{^\prime\mskip-5mu}

\def\ie{i.e.}

\def\msh{{MSH~11$-$6{\sl 2}}}

\def\arund{{$\simeq$}}

\def\la{\hbox{\raise.35ex\rlap{$<$}\lower.6ex\hbox{$\sim$}\ }}
\def\ga{\hbox{\raise.35ex\rlap{$>$}\lower.6ex\hbox{$\sim$}\ }}
 
\received{\underline{October 1997}}
\accepted{\underline{January 1998}}
 
\slugcomment{Accepted for publication in {\it The Astrophysical Journal}}
 
\begin{document}
\title{STUDY OF THE COMPOSITE SUPERNOVA REMNANT MSH~11$-$6{\sl 2} }
 
\author{Ilana~M. Harrus\altaffilmark{1,2,3},  John~P. Hughes\altaffilmark{1,4} and Patrick~O. Slane\altaffilmark{1,5}
\altaffiltext{1}{ Harvard-Smithsonian Center for Astrophysics,
                  60 Garden Street, Cambridge, MA 02138-1516}
\altaffiltext{2}{ Department of Physics, Columbia University,
                  550 West 120$^{\rm th}$ Street,  New York, NY 10027}
\altaffiltext{3}{ imh@head-cfa.harvard.edu}
\altaffiltext{4}{ Department of Physics and Astronomy, Rutgers University,
                  136 Frelinghuysen Road, Piscataway, NJ 08854-8019; 
                  jph@physics.rutgers.edu}
\altaffiltext{5}{ slane@cfa.harvard.edu}}
\begin{abstract}
\par\indent 
We present the analysis of the X-ray data collected during an
observation of the supernova remnant (SNR) \msh\ by the {\it Advanced
Satellite for Cosmology and Astrophysics} (\asca). We show that
\msh\ is a composite remnant whose X-ray emission comes from two
distinct contributions. Nonthermal, synchrotron emission, localized to
a region of radius $\sim$3$^\prime$ (consistent with a point source)
dominates the total flux above 2~keV. A second contribution comes from
a thermal component, extended up to a radius of $\sim$6$^\prime$ and
detected only at energies below 2~keV. The spatial and spectral
analysis imply the presence of a neutron star losing energy at a rate
of about $\dot E \sim 10^{36} - 10^{37}$ ergs s$^{-1}$. No pulsed
emission is detected and we set a limit on the pulsed fraction of
$\sim$10\%.  This is consistent with the lack of a radio pulsar in the
remnant, which may indicate that the pulsed emission from the rapidly
rotating compact object that should be powering the synchrotron nebula
is beamed and our viewing direction is unfavorable. In either event,
the central neutron star deposits much of its spin-down energy into
the surrounding synchrotron nebula where, through direct imaging with
broadband satellites such as \asca, it is possible to study the
energetics and evolution of the compact remnant.
\end{abstract}
 
 
\newpage
%
\section{ Introduction }
In the standard scenario of the core collapse of a massive star, a
shock wave propagates ahead of the ejecta through the interstellar
medium (ISM) forming a supernova remnant (SNR).  The X-ray emission
comes from a thin shell of interstellar material swept-up by the
supernova blast wave and the spectrum shows thermal features
indicative of a high-temperature plasma.  There exists, however, a
class of SNRs which does not follow this standard picture and shows,
instead, a centrally peaked X-ray morphology with a nonthermal
spectrum.  The most famous example of such a remnant is the Crab
Nebula in which the X-ray (and radio) flux is synchrotron radiation
produced by relativistic electrons and magnetic field all powered by
the spin-down energy loss of the compact object, a neutron star,
created in the explosion.  This nonthermal radiation is the signature
of the compact object and provides proof that the remnant was produced
by a Type Ib/c or Type II SN (\ie, the core collapse of a massive star).
In the case of the Crab Nebula, the 33~ms pulsar has been detected at
all wavelengths including the X-ray band.  However, it is often the
case that the pulsed emission itself escapes detection (an effect
which may be due to beaming). X-ray studies provide another tool
to confirm the presence of a compact object by detecting the
associated synchrotron nebula or plerion. This method does not suffer
from the beaming effect bias and gives us direct information on the
energetics, specifically the current energy output, of the central
powering object.  With the advance of the high sensitivity broadband
X-ray imaging and spectral instruments onboard \asca, the study of this
class of objects has made great progress.

\asca\ has made it possible to unambiguously identify and quantify
the nonthermal emission from synchrotron nebulae and to separate it
from the thermal emission seen from shock-heated SNR gas.  The
SNRs for which this distinction has been made have emerged recently
as a growing class. The currently unique imaging capabilities of
\asca\ at energies above \arund~4~keV where the contribution from the
thermal emission is negligible, has allowed us to directly {\it image}
the contribution from the synchrotron emission coming from the
(often undetected) pulsar. \asca\ studies of several SNRs have
either confirmed radio data that suggested the presence of a
synchrotron nebula (e.g., Kes~75, Blanton \& Helfand 1996; and
G11.2$-$0.3, Vasisht et al.\ 1996) or directly identified possible new
members of this class (e.g., W44, Harrus, Hughes \& Helfand 1996;
CTA-1, Slane et al.\ 1997; and MSH~15$-$5{\sl 2}, Tamura et al.\
1996). It is now clear that \msh\ (G291.0$-$0.1) is also a composite
supernova remnant.

Radio observations of \msh\ reveal a centrally-brightened morphology
with a spectral index of $\alpha = 0.29\pm 0.05$ and strong
linear polarization (Roger et al.\ 1986).  Prior to our \asca\ observation,
very little was known about this SNR in the X-ray band.  It was
observed by the \ein\ Imaging Proportional Counter (IPC), but these
data were insufficient to distinguish between a thermal and a
nonthermal interpretation (Wilson 1986). No on-axis pointing of the
object was made with the \rosat\ Position Sensitive Proportional
Counter although an observation with the High Resolution Imager was
recently carried out (but the data have not yet been released to the
public archive).  

Presented here are the results and interpretations of our \asca\ X-ray
data on the SNR \msh.  We present in \S 2 our data extraction methods
and spatial, spectral, and timing analyses. In \S 3 we discuss the
implications of our analysis on the distance to \msh, its evolutionary
state, and the association of the remnant with an unidentified
$\gamma$-ray source.  We interpret the nonthermal emission as having a
synchrotron origin and derive estimates of the mean magnetic field in
the synchrotron nebula that compare well with estimates of the thermal
pressure in the remnant.  Then we tie together the various
observational results into two evolutionary scenarios, both of which
appear to describe the remnant equally well. Further observational
tests that could help distinguish between these scenarios are
discussed briefly. The final section of the article is a summary of
our principal conclusions.

\section{Data Reduction and Analysis}

An \asca\ observation of \msh\ was carried out on 7 February, 1996,
toward the nominal position of the remnant. The event files were
cleaned according to the standard prescription as follows: for the Gas
Imaging Spectrometer (GIS) detectors, we rejected events recorded when
the satellite was traversing regions of low magnetic rigidity ($<$6)
(where the flux of high energy charged particles, and consequently the
X-ray background rate, tends to be greater) or when the Earth elevation
angle was below 10$^\circ$.  
In addition, we also removed the internal background due to the
calibration source at the edge of the detector and discarded data
located on the external bright ring of the detector.  The screening
criteria for the Solid-State Imaging Spectrometer (SIS) data were
almost identical to those used for the GIS: minimum Earth elevation
angle of 10$^\circ$ and a minimum rigidity of 6.  Because of the
possibility for contamination in the SIS of fluorescence lines of
oxygen from the Earth's atmosphere, data selection was also done on
the angle to the limb of the sunlit Earth, and only data above
40$^\circ$ (20$^\circ$) for the SIS0 (SIS1) were retained. Finally,
only events with CCD grades 0, 2, 3, or 4 were used in further
analysis. The various grades characterize the shape of the event as
recorded by the CCD.  Our selection is the standard recommended set
which represents a compromise between the rejection of background
(i.e., charged particle events) and real X-ray photons.

 Our \asca\ observation was planned to study primarily the
spectral nature of the SNR. In particular, we retained GIS rise-time
information in order to pursue accurate spectral analysis and only
assigned two extra bits (from compressing the pulse height spectrum ---
the spectrum is compressed for data analysis in any case) to timing
data.  This gave us, in the GIS, a time resolution of 0.125~sec in
medium-bit-rate mode and 0.015625~sec in high-bit-rate mode.

\subsection{Spatial Analysis}
We generated exposure-corrected, background-subtracted images of the
GIS and SIS data in selected spectral bands. The background was
determined from the weighted average of several nominally blank fields
from high Galactic latitude observations with data selection criteria
matched to those used for the SNR data. Exposure maps were generated from
the off-axis effective-area calibrations, weighted by the appropriate
observation time.  Events from regions of the merged exposure map with
less than 10\% of the maximum exposure were ignored. Merged images of
the source data, background, and exposure were smoothed with a
Gaussian of standard deviation, $\sigma=45^{\prime\prime}$.  We
subtracted smoothed background maps from the data maps and divided by
the corresponding exposure map.  

Fig.~1 shows the strikingly different appearance of \msh\ in the soft,
medium, and hard X-ray bands as observed by the GIS and the SIS.  At
energies below 2~keV (top panels in Fig.~1), the emission is extended,
covering a region with a diameter of somewhat more than 10$^\prime$.
At higher energies, both between 2~keV and 4~keV (medium panels) and
above 4~keV (bottom panels), the extended emission which fills the
radio shell has disappeared and only centrally concentrated emission
appears. Fig.~2 shows the radio contours from the Molonglo Observatory
Synthesis Telescope (MOST) (Whiteoak \& Green 1996) superimposed on
the same X-ray images as displayed in Fig.~1.  There is a clear
correlation between the location of the maximum in radio and the
center of the hard X-ray emission. Furthermore, the total extent of
the radio SNR agrees fairly well with that of the soft X-ray contours
and we estimate the angular radius of the remnant based on the
outermost radio contours in Fig.~2 to be $6\myarcmin.4$.

In Fig.~3 we show the variation of an X-ray hardness ratio (which we
define to be the number of counts between 2 and 10 keV divided by the
number of counts between 0.5 and 2 keV) as a function of radius.  The
ratio decreases significantly with radius (which is opposite to the
sense of instrumental effects) and confirms that the central emission
is considerably harder than the emission from the edge.  We show below
through our spectral analysis that the central hard emission is
dominated by a nonthermal, power-law contribution.  The size of the
hard X-ray source is consistent with an unresolved source
($<$3$^\prime$ radius in both SIS and GIS) and its location is
(11$^{\rm h}$11$^{\rm m}$48.5$^{\rm s}\pm$1.0$^{\rm s}$,
$-$60$^\circ$39$^\prime$20$^{\prime\prime}\pm$15$^{\prime\prime}$) in
the GIS and (11$^{\rm h}$11$^{\rm m}$48.7$^{\rm s}\pm$1.0$^{\rm s}$,
$-$60$^\circ$40$^\prime$00$^{\prime\prime}\pm$15$^{\prime\prime}$) in
the SIS. This position is consistent with the highly polarized, flat
spectrum region at the center of the radio image of \msh.  We estimate
the angular size of this region to be $2\myarcmin.5$ in radius, where, to
account for the significant elongation of the nebula, we have taken
the geometric mean of the semi-major and semi-minor axis lengths of
the region defined by the third lowest radio contour in Fig.~2.
\par 
Above 2.0~keV, the GIS and SIS images reveal a hard extended source
located at the northeastern edge of the field of view (at approximate
position 11$^{\rm h}$13$^{\rm m}$40$^{\rm s}$,
$-$60$^\circ$33$^\prime$30$^{\prime\prime}$) that is absent at low 
energies.  This source is likely to be the ghost image of the nearby
bright X-ray binary Cen X-3 which is located about $1^\circ$ east of
\msh.
Ghost images are created when X-rays from a source outside the field
of view are scattered off only one of the focusing mirror elements.
As Fig.~1 shows, this spurious emission is isolated from the center of
the field of view where \msh\ lies and, thus, has no impact on our
analysis.

\subsection{Spectral Analysis}

We extracted X-ray spectra from circular regions centered at 
11$^{\rm h}$11$^{\rm m}$51.49$^{\rm s}$, $-$60$^\circ$39$^\prime$11.74$^{\prime\prime}$ (J2000), using a radius of 5$^\prime$ in the GIS and, due to its
smaller field of view, only 4$^\prime$ in the SIS.  After the standard
data selection criteria were applied and within the regions described
above, there were 4760~(4949) events in GIS2 (GIS3) and 4726~(4182)
events in SIS0 (SIS1).  We first fitted the SIS and GIS separately and
then jointly. The extended nature of the source and the contamination
by the ghost image of Cen X-3 made it impossible to extract a
sufficient amount of background data from a blank part of the field,
especially for the SIS analysis. For background, therefore, we 
used the high Galactic latitude blank fields available from the {\it
HEASARC}, after establishing with the GIS data that the high-energy
contamination from the Galactic plane was negligible.

\par 

To take into account the high-energy spectrum as well as the extended
soft component present below 2~keV, we modeled the data with a
mixture of nonthermal (\ie, power-law) and thermal components.  For
the thermal emission we used the Raymond \& Smith (1977) thermal plasma
model with cosmic abundances (Anders \& Grevesse 1989). Absorption
along the line of sight was taken into account with an equivalent
column density of hydrogen, $N_{\rm H}$, using the cross-sections and
abundances from Morrison \& McCammon (1983). A
purely nonthermal model failed to take into account the low energy
part of the spectrum, while a single thermal model failed to account
for the high energy part. Considerably better fits in each case were
obtained by including an additional thermal component. However, the
exclusively thermal model required an unseemly high second temperature
($\sim$12~keV) to account for the high-energy emission.  The best
overall fit ($\chi^2 = 671.9$, $\chi^2_{\rm r}\sim 2.0$) to the
joint GIS and SIS spectral data was obtained for a model including
both thermal and nonthermal components.  The power-law component has
a photon index of $2.0^{+0.1}_{-0.3}$ (the errors quoted here take
into account the different analyses carried out --- detectors fitted
separately, simultaneously, and combined with their respective
response file --- as well as the standard 1 $\sigma$ statistical error).
Each pair of nominally identical detectors (GIS and SIS) produced
consistent normalizations and so they were linked in the remaining
fits.  We find the flux density of the power-law at 1~keV in the GIS
to be $F_{\rm 1\, keV} = (1.26\pm0.13)\times10^{-3}$
photons~cm$^{-2}$~s$^{-1}$~keV$^{-1}$ and $F_{\rm 1\, keV} 
= (1.0\pm0.1)\times10^{-3}$ photons~cm$^{-2}$~s$^{-1}$~keV$^{-1}$ in
the SIS.  (N.B., the GIS and SIS normalizations differ because the
X-ray flux that is lost in the gaps between the CCD chips in the SIS
is not accounted for in the effective area used to model these data.)
The thermal component has a temperature $kT = 0.80\pm 0.05$ keV and an
emission measure ($n_{\rm H}n_eV / 4\pi D^2$) of
$(1.9^{+1.0}_{-0.5})\times 10^{11}$ cm$^{-5}$ 
from the GIS and $(8.6^{+5.7}_{-2.8})\times 10^{10}$ cm$^{-5}$ from the
SIS. 

The column density toward the source is $N_{\rm H} = (6\pm1)\times
10^{21}$ atoms~cm$^{-2}$, consistent with the previously derived value
of $4\times10^{21}$ atoms~cm$^{-2}$ (Wilson 1986).  The unabsorbed
flux from this thermal component between 0.1 and 2.0~keV, computed
with the GIS normalization because of its more complete coverage of
the source, is $(6.3^{+4}_{-2})\times 10^{-12}$
ergs~cm$^{-2}$~s$^{-1}$ contributing 51\% of the total flux in this
energy range, while it is only $(2.4^{+1.8}_{-0.9})\times 10^{-13}$
ergs~cm$^{-2}$~s$^{-1}$ above 2.0~keV, less than 7\% of the total flux
of $(3.5^{+1.1}_{-0.8})\times 10^{-12}$ ergs~cm$^{-2}$~s$^{-1}$ from
the remnant.

Fig.~4 shows the data and the best-fit power-law plus thermal
model. The thermal component is plotted separately in the insert to
the figure. Table~1 gives numerical results from the spectral
analysis.  The residuals in Fig.~4 show the presence of an additional
line emission that is not properly taken into account in the standard
plasma model we used to fit the data. We added a line at 1.3~keV
(roughly where the Mg K$\alpha$ line should be) to the model, fixed
the energy and the width of the gaussian profile, and fitted for the
intensity of the line. We found that $\chi^2$ dropped by more than 30
when the line was included. The best-fit equivalent width (derived
from the SIS) was 40~eV.  This extra Mg emission could be the result
of nonequilibrium ionization effects or enhanced abundances, but the
statistical precision of the \asca\ data is insufficient to allow us
to give a definitive explanation at this time. Nevertheless, these
clear indications of thermal emission, combined with our spatial
study, allow us to conclude that we are seeing the two components
predicted by the standard model of SNR evolution: soft shell-like
emission (temperature \arund~0.8~keV) from the interaction of the
shock wave with the interstellar medium and the hard nonthermal
contribution arising from synchrotron emission of relativistic
electrons in the magnetic field of the nebula surrounding the neutron
star.

\subsection{Timing Analysis} 

As emphasized in the previous section, the presence of a compact
object in \msh\ is unambiguous and so one might expect to detect the
pulsed signal coming from the the neutron star itself. In fact, not
every existing pulsar is expected to be detected; beaming effects, for
example, may result in a cone of emission which does not intersect the
Earth. This beaming effect presumably accounts for the relatively low
number of SNRs in which a radio pulsar, undoubtedly associated with
the remnant, has been discovered. In fact, of the 24 cataloged SNRs
(Green 1996) that show some evidence for a nonthermal X-ray spectrum,
only half have been associated with radio pulsars.  In addition, a
recent paper (Kaspi et al.\ 1996) presented the results of a search for
radio pulsars in a sample of 40 southern Galactic SNRs.  \msh\ was
part of the sample investigated but no pulsar was detected in the
vicinity of the SNR. The limits on the radio flux density deduced from
this unsuccessful search are shown on Fig.~5.

Because our primary emphasis was to acquire accurate spectral data,
the observational configuration of our data set was not optimized for
timing analysis.  Moreover the statistics are low: after further
restricting ourselves to the high-energy ($>$4 keV), central portion
of the remnant there were only $\sim$1230~events collected at a time
resolution of 125~ms and even fewer ($<$300) in high-bit-rate mode
(resolution $\sim$16~ms). Both FFT and $Z_{\rm n}$ tests for the
125~ms resolution data found some possible periodicities (at about the
3~$\sigma$ level), suggestive of a pulsed fraction of $\sim$10\%, but
the results were not confirmed in the high-bit-rate data and remain
unconvincing. If there was a pulsar and it was 10\%-pulsed, then we
would expect to see a point source in the IPC image with an average
count rate of $6.0^{+2.4}_{-1.3}\times 10^{-3}$ cts~s$^{-1}$ assuming the
\asca-derived parameters of the power-law spectral component.  In fact
there is no central point source in the IPC image with a count rate this
large, effectively ruling out a pulse fraction of 10\% or greater.

\section{ Discussion}
\subsection{Distance to \msh}
As is often the case in astronomy, the distance to \msh\ is not well
known. No measure of H{\sc i} absorption is available for this source and,
thus, the method sometimes used in this situation correlates the
remnant's surface brightness, $\Sigma$, (a distance independent
quantity) to its diameter, $D$. This is the $\Sigma$--$D$ relation in
which, for a sample of SNRs, one fits a power-law function to the
variation of $\Sigma$ with $D$. This method has been applied to \msh\
(Clark \& Caswell 1976; Milne 1979) and leads to a distance of 7--11
kpc, where the range comes from the sample of SNRs on which the
relation is defined.  The $\Sigma$--$D$ relation is, however, known to
be unreliable, especially for centrally-brightened SNRs like \msh.  A
better estimate of the distance, for Crab-like SNRs, can be derived
using the $S\theta^2$-$D$ relation (Weiler \& Panagia 1980). As
opposed to the shell-like SNRs used in the $\Sigma$--$D$ relation,
only plerions are used in this method that links the flux density
($S$) and angular size ($\theta$) to the distance ($D$).  Using this
method, the distance to \msh\ is found to be about 6~kpc.  Yet
another estimate of the distance to \msh\ can be found using a
$\Sigma$--$D$ relation by Reynolds \& Chevalier (1984) adjusted to
model the Crab very well. This gives a small distance (between 1 and
6~kpc).  Finally, Roger et al.\ (1986) suggest a distance to \msh\
similar to that of two H{\sc ii} regions (G290.6$+$0.3 and
G291.3$-$0.7) present in the vicinity of the SNR. This hypothesis
leads to a distance of 3.5~kpc, in agreement with the bounds found in
the Reynolds \& Chevalier model.

\par 

Another method to estimate the distance uses the correlation between
column density (absorption along the line of sight) and distance (see
Fig.~3 in Seward et al.\ 1972).  For a column density of
$6\times10^{21}$ atoms~cm$^{-2}$, this method gives an estimate of the
distance between 2 and 6~kpc.  One should note that this method makes
use of an average density along the line of sight, an approximation
which certainly introduces a considerable uncertainty in the distance
estimate. As we discuss below, one interpretation of the X-ray
results is suggestive of a larger distance than those mentioned
previously. In our analysis, we have kept the distance dependence
explicit and express all our results as a function of $D_{\rm 10}$
(where $D_{\rm 10}$ is the distance normalized to 10 kpc.)

\subsection{Dynamical Evolution of \msh}
From the size of the remnant, the observed emission measure, and the
distance, one can estimate the density of the X-ray--emitting gas in
\msh. The volume of the emission region defined by the total extent of
the SNR (an angular radius of $6\myarcmin.4$) assuming spherical
symmetry is $V = 7.9\times
10^{59} \,f\,D_{10}^{3}$ cm$^{-3}$, where $f$ is the volume filling
factor of the hot X-ray--emitting gas, which is less than
unity. Combined with the observed 
emission measure determined from the spectral analysis (Table 1) and a
ratio of $n_e/n_{\rm H}=1.17$, we deduce a hydrogen number density of
$n_{\rm H} = (5.0^{+1.1}_{-0.7})\times 10^{-2}\,D_{\rm
10}^{-1/2}\,f^{-1/2}$ cm$^{-3}$.  This is the mean density of the
shocked gas and is independent of the dynamical state of the remnant.
We study the dynamical evolution of \msh\ assuming two extreme values
for the remnant's distance.  The short distance, 3.5~kpc, is
consistent with the distance determined by the X-ray column density,
the H{\sc ii} regions, and the Reynolds \& Chevalier $\Sigma$--$D$
relation. For this distance, we find a mean density in the remnant of
$n_{\rm H} \sim 8.5\times 10^{-2}\,f^{-1/2}$ cm$^{-3}$ and a 
mass of X-ray emitting plasma of $M_{X} = 3.2\,f^{1/2}$ $M_\odot$
(assuming a helium abundance 
0.085 times hydrogen by number).  Given this small amount of 
mass, the remnant clearly cannot be in the Sedov-Taylor phase of
evolution (Taylor 1950; Sedov 1959) and should rather be considerably
younger. One limit on the age is set by assuming undecelerated
expansion of the remnant at 5000 km s$^{-1}$, which gives a free
expansion age of 1300 yr. (The remnant is 6.5 pc in radius at this
distance.) 
A firm upper limit of 3500 yr is provided by (erroneously!) assuming
Sedov-Taylor expansion (see below). If this scenario is correct, then
it is perhaps surprising that the thermal emission from \msh\ does not
show stronger evidence for enhanced metal abundances, which might be
expected from the presence of reverse-shock heated ejecta.  The low
density of the ambient medium would, however, mitigate against a
strong reverse shock.  A similar absence of an ejecta component is
also observed for the Crab Nebula and 3C58, both of which reside in
low density environments.  Finally, we note that the mean thermal
pressure inside the SNR is 
$P = n kT \sim 2.5\times 10^{-10}\,f^{-1/2}$ dyne cm$^{-2}$, where $n$ is 
the total number density of particles (electrons and ions) and the temperature, $kT$, is the same for all species.

In view of the large uncertainties that plague our estimates of the
distance to \msh, we have examined the possibility that the remnant is
located at a much larger distance, 10 kpc.  In this case the hot gas
mass is considerable, $M_{X} = 44\,f^{1/2}$ $M_\odot$, and so it is
reasonable to assume that the remnant is in the Sedov-Taylor or
adiabatic phase of evolution. The self-similar solutions to the
hydrodynamical evolution of adiabatic-phase SNRs provide a numerical
relationship between the age of the remnant and the size and
temperature of the shock wave. The temperature determined by our
spectral analysis (\S 2.2), $\langle T\rangle$, is the
emission-measure-weighted average electron temperature, which is
proportional to the shock temperature, $\langle T\rangle = 1.27 T_S$
(assuming, as above, that electrons and ions have the same
temperature).  The Sedov age relationship is $t\approx 480\,{\rm
yrs}\,(R /1\,{\rm pc})\,(\langle T\rangle /1\,{\rm keV})^{-1/2}$, and
so the measured angular size and temperature of \msh\ imply an age of
\arund\,$10^4\,D_{10}$ yr. We estimate the preshock Hydrogen number
density $n_0$ by integrating the interior radial density variation
from the Sedov similarity solution and then equating it to the
observed emission measure. This gives a preshock density of $n_0 =
(3.4\pm0.7) \times 10^{-2}\, D_{10}^{-1/2}$ cm$^{-3}$.  With the
preceding numerical values we estimate the supernova explosion energy
from the Sedov relations to be $E = (0.25\pm0.05) \times
10^{51}\,D_{10}^{5/2}$ ergs.  The central pressure $P_0$ in the Sedov
similarity solution is related to the pressure at the shock front as
$P_0 = 0.31 P_S$.  Evaluting using the best fit numerical quantities
we find $P_0 \sim 9.6\times 10^{-11}$ dyne cm$^{-2}$.

\par
\subsection{\msh\ and the {\it EGRET} Source 2EG J1103$-$6106: A Doubtful Association}

The Energetic Gamma-Ray Experiment Telescope ({\it EGRET}) on board
the Gamma Ray Observatory detected a source in the general vicinity of
\msh\ with an integrated flux from 100~MeV to 30~GeV of
$(5.6\pm0.9)\times10^{-7}$ photons~cm$^{-2}$~s$^{-1}$. The second
\egret\ catalog (Thompson et al.\ 1995) lists the source as 2EG
J1103$-$6106 and gives a position of 11$^{\rm h}$03$^{\rm m}$48$^{\rm
s}$, $-$61$^\circ$06$^\prime$ (J2000) with an error ellipse of size
$49^\prime \times 32^\prime$ that is only marginally consistent with
the position of \msh.  For this source, the \egret\ team (Merck et
al.\ 1996) reports a spectral index of $-2.3\pm 0.2$ (marginally consistent
with the X-ray spectral index reported here). Fig.~5 shows the {\it
EGRET} spectrum of 2EG J1103$-$6106 in comparison to data on \msh\
from radio to X-ray wavebands. The $\gamma$-ray data are inconsistent
with a simple extrapolation of the X-ray power-law spectrum to higher
energies. Note that the total number of photons detected by \egret\ is
small, only 312, and that the position is uncertain by more than
30$^\prime$. Moreover, a recent paper (Kaspi et al.\ 1997) reported
the discovery of a pulsar, possibly associated with the nearby SNR
MSH~11$-$6{\sl 1}A, at a position that agrees with the error box of
2EG J1103$-$6106. These authors conclude that the $\gamma$-ray source
and this pulsar were likely to be associated. Our spectral results are
consistent with this interpretation (i.e., that the {\it EGRET} source
and \msh\ are not associated) and so we exclude the {\it EGRET}
$\gamma$-ray source from further discussion.

\par
\subsection{Empirical Models for the Synchrotron Nebula}
We have argued above that the nonthermal emission is coming from 
the synchrotron nebula surrounding a spinning neutron star, or
pulsar, created in the supernova explosion. From the empirical
relationship between a pulsar's spin-down energy $\dot E$ and the
power-law X-ray luminosity of the nebula $L_{X}$ (Seward \& Wang
1988), which is $(2.5\pm 0.3)\times 10^{34}\,D_{10}^2$ ergs s$^{-1}$
(0.2 -- 4.0 keV), we infer $\dot E\sim (1.0 \pm 0.9) \times
10 ^{37}\, D_{\rm 10}^{1.44}$ ergs s$^{-1}$ for the putative pulsar in
\msh. 

In figure 5 we plot the available radio spectral data (Roger et
al.~1986) and the results of the power-law fits for the X-ray
synchrotron nebula.  For the following discussion the points on the
figure indicating the upper limits to the flux of the radio pulsar
(downward-pointing arrows on the left side) and the \egret\
$\gamma$-ray source (plus symbols on the right side) should be
ignored. We can use the radio spectral data, the well-established
theory for synchrotron radiation, and a minimum total energy (in
fields and particles) assumption to estimate the magnetic field
strength in the nebula.  For a power-law distribution of electrons,
one can write (Ginzburg \& Syrovatskii, 1965, eqn.~5.16):
\begin{equation}
B_{\rm 1} = \left [ 48\,\kappa_{\rm m}\,\kappa_{\rm r}\,A(\gamma,\nu)\frac{F_{\nu}}{D\, \theta^{\rm 3}} \right ]^{\rm 2/7} 
\end{equation}
where $\kappa_{\rm m}$ is the constant of proportionality between the
energy in magnetic field and relativistic particles (minimum total
energy is obtained for $\kappa_{\rm m} =3/4$, which is the value we
use) and $\kappa_{\rm r}$ is the proportionality constant between the
energy in relativistic electrons and the energy in cosmic rays of all
species. If the protons and electrons are assumed to have the same
energy density throughout the source, then we have $\kappa_{\rm r}$ =
2 which is the value adopted in the following computation.
$A(\gamma,\nu)$ is a numerical coefficient (which is calculated from
the physics of synchrotron radiation) that depends on the slope of the
electron energy spectrum ($\gamma$) and the frequency range over which
the power-law distribution is valid.  In the radio band $\gamma =
2\alpha +1$, where $\alpha$ is the radio spectral index (defined as 
$F_\nu \sim \nu^{-\alpha}$). The quantities $D$ and $\theta$ are
respectively the distance to the object and 
its angular diameter.  Note that the frequency dependence of the quantity
$A(\gamma,\nu)$ is cancelled by the corresponding $\nu$ dependence of
the spectral flux density of the source, $F_{\nu}$.  One can write
(explicitly including the distance dependence):
\begin{equation}
B_1 = 46.9\, D_{10}^{-2/7}\, (\theta_{\rm P}/2\myarcmin.5)^{-6/7}
\left [ \kappa_{\rm m}\,\kappa_{\rm r}\right ]^{2/7}\mu{\rm G}.
\end{equation}
For the size of the radio plerion we derived an average angular radius
$\theta_{\rm P} = 2\myarcmin.5$ 
from the MOST radio image as discussed above. 
Table 2 summarizes the values of $B_{\rm 1}$ found for the two different
distances considered. 

The broadband spectrum of the nebula in \msh\ shows a clear spectral
break somewhere between the radio and the X-ray band.  The break
frequency $\nu_{\rm B}$, which is where the extrapolated X-ray and
radio power-law spectra intersect, lies in the range 8--200 GHz. The
lower limit corresponds to not violating any of the radio
measurements, while the upper limit is derived from the largest
allowed slope of the X-ray power-law spectrum.  Assuming that
$\nu_{\rm B}$ represents the frequency at which synchrotron losses
begin to dominate, we can obtain another estimate of the nebular
magnetic field. In this case, we find
\begin{equation}
B_2 = 740\,(\nu_{\rm B}/35~{\rm GHz})^{-1/3}\,(t/10^4~{\rm
yrs})^{-2/3} \,\mu{\rm G}.
\end{equation}
which comes directly from Ginzburg \& Syrovatskii (1965,
eqn.~5.36). Here $t$ is the age of the nebula, for which we use the
values derived in our study of the thermal emission (tabulated in
Table 2).  This estimate of the magnetic field strength has a large
uncertainty because of the large range in break frequencies.
Furthermore, this method is to be considered with caution. The break
frequency deduced from extrapolating the X-ray and radio spectra is
not a well defined physical quantity. In fact, there are other
remnants [e.g. 3C58 (Green \& Scheuer 1992)] for which no simple
extrapolation from X-ray to radio spectra is possible, suggesting
other evolutionary effects for the central pulsar. In such cases,
computation of the magnetic field with this method is no longer valid.

Yet a third method to compute the magnetic field supposes pressure
equilibrium between the hot thermal gas in the center of the SNR and
the synchrotron nebula so that $B_3 = \sqrt{8 \pi P_0}$.  We obtained
estimates of the pressure in the hot gas in \S 3.2 for both assumed
distances to the remnant (values are shown in Table 2) which result in
values for $B_3$ of 79 $\mu$G and 50 $\mu$G. These values are
potentially rather uncertain as well. The most significant uncertainty
arises from our assumption during the analysis of the thermal emission
that the electron and ion temperatures in the remnant were equal.  It
is possible and perhaps likely that the electrons have significantly
lower temperatures than the ions because the timescale for exchange of
energy between these different species is quite long if only Coloumb
collisions operate. In fact the protons could have 10 to 100 times the
temperature of the electrons and that would increase our magnetic
field estimates by factors of 3 to 10.

\subsection{ Interpretation}
The two different distance regimes discussed above lead to distinctly
different evolutionary stages for MSH~11--6{\sl 2}. At a distance of
3.5~kpc, the remnant is dynamically young, having swept up only a
small amount of mass. 
The ambient density $n_0 \sim 0.04 {\rm\ cm}^{-3}$ (assuming a volume 
filling factor of $f = 1/4$ and the strong shock compression factor of 4)
is quite low, possibly indicating evolution in a
wind-blown cavity created by the massive precursor star implied by the
presence of the compact remnant. The luminosity of the synchrotron nebula is
$L_{synch} = 3.1 \times 10^{33}{\rm\ erg\ s}^{-1}$, suggesting a
central pulsar with a spin-down energy loss rate of $\dot E = 2.5
\times 10^{36} {\rm\ erg\ s}^{-1}$. For a surface magnetic field of
$B_s = 10^{12}-10^{13}$~G and a braking index $\nu = 2.5-3$, this
implies an initial spin period in the range $P_0 \sim 45-130$~ms with
a current-epoch period in the range $P_0 \sim 46-160$~ms. The magnetic
pressure associated with the synchrotron nebula is $B_{neb} \sim 70
\mu$G and is in close equilibrium with the thermal pressure of the SNR
interior.

For a distance as large as 10~kpc, \msh\ would be in the Taylor-Sedov phase
of evolution. The swept-up mass  $\approx 30 M_\odot$ 
is sufficiently large to explain the lack of a significant ejecta component
to the thermal emission, and the associated explosion energy $E_{51} \approx
0.25$ is rather typical of SNRs. The remnant is then $\sim 10^4$~yr old and
harbors a synchrotron nebula whose luminosity is $L_{synch} = 2.5 \times 
10^{34}{\rm\ erg\ s}^{-1}$. The associated spin-down power 
is $\dot E = 10^{37}{\rm\ erg\ s}^{-1}$ and the neutron star parameters 
above yield an initial
spin period range of $29-46$~ms with a current-epoch period in the range
of $31-100$~ms. (We note that for some values of $P_0$ in the above
range, $P$ and $B_s$ are double-valued, corresponding to solutions for $P
\approx P_0$ and $P$ significantly larger than $P_0$.)
These parameters are similar to other young pulsars (e.g.
PSR~B1509--58 in MSH~15$-$5{\sl 2}, or PSR~B1800--21 in G8.7--0.1) making this scenario
plausible. However, the ambient density 
$n_0 \sim 3.4 \times 10^{-2} {\rm\ cm}^{-3}$ is rather small. One might again invoke a pre-SN
wind-blown cavity as an explanation, but in this case the size of the
nebula would need to be considerable $\sim$ 20 pc in radius.
Local variations in density by factors as large as 5 or more are
observed in the direction of \msh\ (Burton \& te Lintel Hekkert 1986),
but the integrated column density is also problematic. The total column
density integrated through the entire galaxy in the direction
toward \msh\ is $N_H = 1.5 \times 10^{22}{\rm\ cm}^{-2}$ (Dickey
\& Lockman 1990) while that for $D > 10$~kpc is only $\sim 3 \times 10^{21}
{\rm\ cm}^{-2}$ (obtained by rough integration of the H{\sc i} gas 
density maps of the outer Galaxy from
Burton \& te Lintel Hekkert). It would thus appear difficult to place the
remnant at 10~kpc given the column density derived from the spectral fits.
We note, however, that the H{\sc i} values are averaged over a grid size larger
than the size scale of the remnant, so that fluctuations corresponding to
the observed X-ray column density over a distance of 10~kpc cannot be
completely ruled out. 

H{\sc i} absorption measurements for the SNR would lead to an accurate 
determination 
of the distance which would resolve the ambiguity in determination of the
evolutionary state of the remnant. If \msh\ is indeed older and more distant,
then one might expect the blast wave to have encountered a number of clouds
in which the shocks have become complete. Thus, optical observation of
H$_\alpha$ and [O{\sc iii}] or [S{\sc ii}] would be of interest in
attempting to differentiate between the two scenarios.

\section{Summary}
We have presented the results of \asca\ X-ray spectral and spatial
studies of the SNR \msh. This remnant is now clearly identified as a
composite SNR with an unresolved central synchrotron-emitting
component powered by a compact object and an extended thermal
component of ISM plasma shock-heated by the SN blast wave. No pulsed
signal from the unresolved central component is detected in the X-ray
band, consistent with the failure of previous radio searches to detect
a radio pulsar. We find that we cannot confidently determine the
evolutionary state of the remnant; the main problem is the lack of a
good distance estimate.  If the remnant is nearby (3.5 kpc), it is
young ($<$ 3500 yr), has not reached the Sedov phase, and is evolving
in a very low density environment. If the remnant is further away, say
10 kpc, then it would be a middle-aged remnant in its
Sedov phase of evolution, with an initial explosion energy of $\sim 2
\times$~10$^{\rm 50}$~ergs.  For either distance we can find a set of 
parameters that can plausibly describe the central pulsar now
powering the synchrotron nebula. Likewise we can develop an
acceptable description for the synchrotron nebula that is consistent
with those of other known synchrotron nebulae for which the powering neutron
star has been detected by radio or X-ray timing searches. Our results
confirm the power of direct imaging and spectral analysis of X-ray
data for detecting compact objects inaccessible to pulse searches.  It
does not fall prey to beaming effects and this technique should be
used more extensively to survey all the cataloged SNRs that have, as
does \msh, a flat and highly polarized radio spectrum.

\par 
David Helfand and Philip Kaaret carefully read drafts of this paper
and we thank them for their many comments that improved it. We thank
Bryan Gaensler for providing the MOST data in electronic form.  This
research was supported in part by NASA under grants NAG 5-3486 and
contract NAS 8-39073. JPH acknowledges support from NASA Grants NAG
5-4871 and NAG 5-4794.

\pagebreak

\clearpage
\begin{figure}
\vspace{-1.0in}
\caption{\asca\ X-ray images of the SNR \msh\ in three X-ray bands (top: 0.5--2.0~keV; middle: 2.0--4.0~keV; bottom: \ga 4.0~keV) from the GIS (left) and the SIS (right). Contour values are linearly spaced from 30\% to 90\% of the peak surface brightness in each map. Peak/background values are top: 1.4/0.084; middle: 1.3/0.042; bottom: 0.81/0.053 for the GIS and top: 2.5/0.145; middle: 1.5/0.042; bottom: 0.71/0.053 for the SIS, where all values are quoted in units of $10^{-3}$counts~s$^{-1}$~arcmin$^{-2}$.}
\end{figure}
\begin{figure}
\vspace{-1.0in}
\caption{Same as Fig.~1 with the radio contours from the MOST survey superimposed. The radio contours are linearly spaced from a minimum value of 0.027 Jy to a maximum of 0.27 Jy.} 
\end{figure}
\begin{figure}
\vspace{-1.0in}
\caption{Variation of the X-ray hardness ratio as a function of radius for \msh. Both GIS (with detectors 2 and 3 combined) and SIS (with detectors 0 and 1 combined) profiles are shown.} 
\end{figure}
\begin{figure}
\vspace{-1.0in}
\caption{GIS and SIS spectra of \msh\ extracted from circular regions centered on the remnant. There were 9709 GIS events within a circle of radius 5$^\prime$ and 8908 SIS events within a circle of radius 4$^\prime$. The solid curves in the top panels show the joint best-fit Raymond \& Smith thermal plasma plus power-law models. The insert figures show only the thermal-component; above 2.0~keV, its contribution is an order of magnitude less than the power-law emission.  The bottom panels plot the data/model residuals. There appears to be excess emission near 1.3 keV, which can be modeled well as a K$\alpha$ line of helium-like magnesium (see text).}
\end{figure}
\clearpage
\begin{figure}
\vspace{-2.0in}
\caption{Broadband plot of the spectral flux density (in Jy) vs.\ frequency for \msh. The radio points correspond to frequencies of 0.408~GHz, 1~GHz, 5~GHz, and 8.4~GHz (Roger et al.\ 1986) and the line through them denotes the best radio spectral index (dotted when extrapolated beyond the frequency range of the measurements). Also indicated (as downward-pointing arrows near the left side of the figure) are the upper limits on a pulsed component to the radio flux (Kaspi et al.\ 1996). In the X-ray band ($\sim$10$^{17}$ Hz), the \asca\ spectral results for the power-law component are shown as two curves bounding the allowed range of X-ray spectral indices. Extrapolations of the X-ray spectra beyond the \asca\ band (0.5--10 keV) are shown dashed. The $\gamma$-ray spectrum of the \egret\ source 2EG J1103$-$6106 (Merck et al.\ 1996) is shown, although the association of this source with \msh\ is unlikely (see text).}
\end{figure}
\clearpage

\footnotesize
\centerline{\bf Table 1}
\centerline{\bf Spectral Analysis of Power Law plus Thermal$^{\rm\  a}$
Plasma Model}
\vspace{0.5cm}
\centerline{\begin{tabular}{lc} \tableline\tableline \\[-8mm]
\multicolumn{1}{l}{Parameter} &
\multicolumn{1}{c}{Fit results$^{\rm b}$} \\[1.5mm] \tableline 
$N_{\rm H}$ (atoms~cm$^{-2}$)& (0.6$\pm$0.1)$\times$10$^{\rm 22}$\\[1.5mm]
$\alpha_{\rm p}$ &  2.0$^{+0.1}_{-0.3}$  \\[1.5mm] 
$F_{\rm 1 keV}$ (photons cm$^{\rm -2}$ s$^{\rm -1}$ keV$^{\rm -1}$)& (1.26$\pm$0.13)$\times$10$^{\rm -3}$\\[1.5mm]
$kT$ (keV) &   0.80$\pm$0.05 \\[1.5mm]
EM$^{\rm c}$ (cm$^{\rm -5}$) & (1.9$^{\rm +1.0}_{ \rm -0.5}$)$\times$10$^{\rm 11}$    \\[1.5mm] 
$\chi^{2}$/$\nu$ & 671.9/336  \\[1.5mm] \tableline
\multicolumn{2}{l} {$^{\rm a}$ Cosmic abundances (Anders \& Grevesse
1989)}\\
\multicolumn{2}{l} {$^{\rm b}$ Single-parameter 1~$\sigma$ errors}\\
\multicolumn{2}{l} {$^{\rm c}$ Emission measure (${n_{\rm H}n_eV \over 4\pi D^2}$)}\\
\end{tabular}}

\newpage
\centerline{\bf Table 2}
\centerline{\bf Parameters derived for MSH~11$-$62 }
\vspace{0.2cm} 
 \centerline{\begin{tabular}{lcc} \tableline\tableline \\[-0.8cm]
 & \multicolumn{2}{c}{ Distance (kpc) } \\[1.5mm] \cline{2-3}
Parameter                         & 3.5 & 10 \\[1.5mm] \tableline
$R_{\rm S}^{\rm\ a}$ (pc)     & 6.1 & 18.6  \\[1.mm] 
Age (yr)                          & $\sim$2500 & $\sim$$10^4$ \\ [1.mm]
$R_{\rm P}^{\rm\ b}$ (pc)   & 2.5 & 7.3  \\[1.mm] 
$P_0$$^{\rm\  c}$ ($10^{-10}$ dyne cm$^{-2}$) & $2.5 f^{-1/2}$ & 1.0 \\[1.mm]
%
%
$B_{\rm 1}$ ($\mu$G)& 72 & 53 \\ [1.mm]
$B_{\rm 2}$$^{\rm\  d}$ ($\mu$G)& 160--500 & 400--1200 \\ [1.mm]
$B_{\rm 3}$$^{\rm\  c}$ ($\mu$G)& $79f^{-1/4}$ & 50 \\[1.5mm] \tableline 
\multicolumn{3}{l} {$^{\rm a}$ Angular size of shock radius is
$6\myarcmin.4$}\\ 
\multicolumn{3}{l} {$^{\rm b}$ Angular size of radio plerion is
$2\myarcmin.5$}\\ 
\multicolumn{3}{l} {$^{\rm c}$ $f$ is the volume filling factor}\\ 
\multicolumn{3}{l} {$^{\rm d}$ Range quoted corresponds to break
frequencies between 8--200 GHz}\\
\end{tabular}}

\newpage
\pagestyle{empty}
\normalsize
\vspace{-.05in}
\begin{figure}[h]
\vspace{-1.0in}
\centerline{\psfig{file=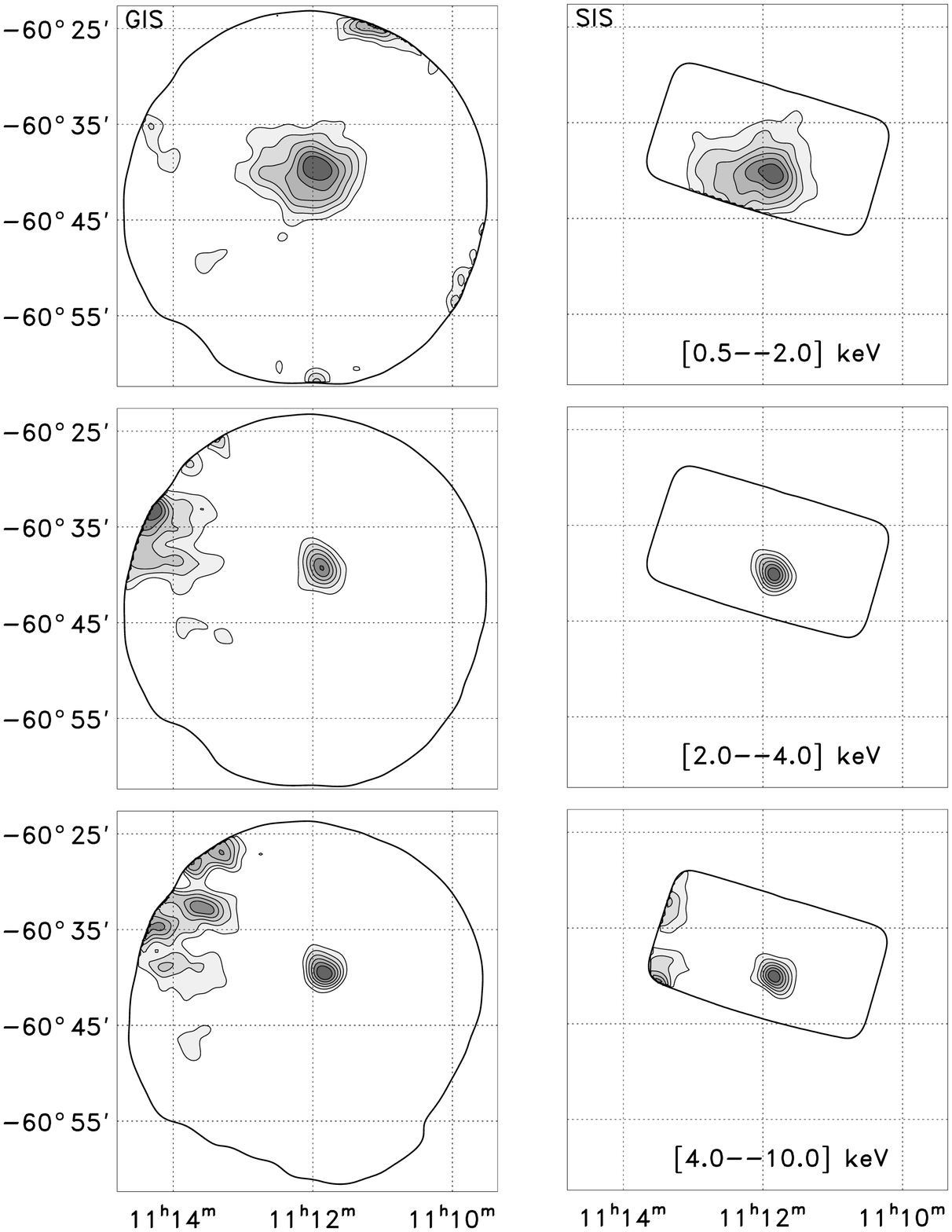,width=6in}}
\end{figure}

\begin{figure}[t]
\vspace{-1.0in}
\centerline{\psfig{file=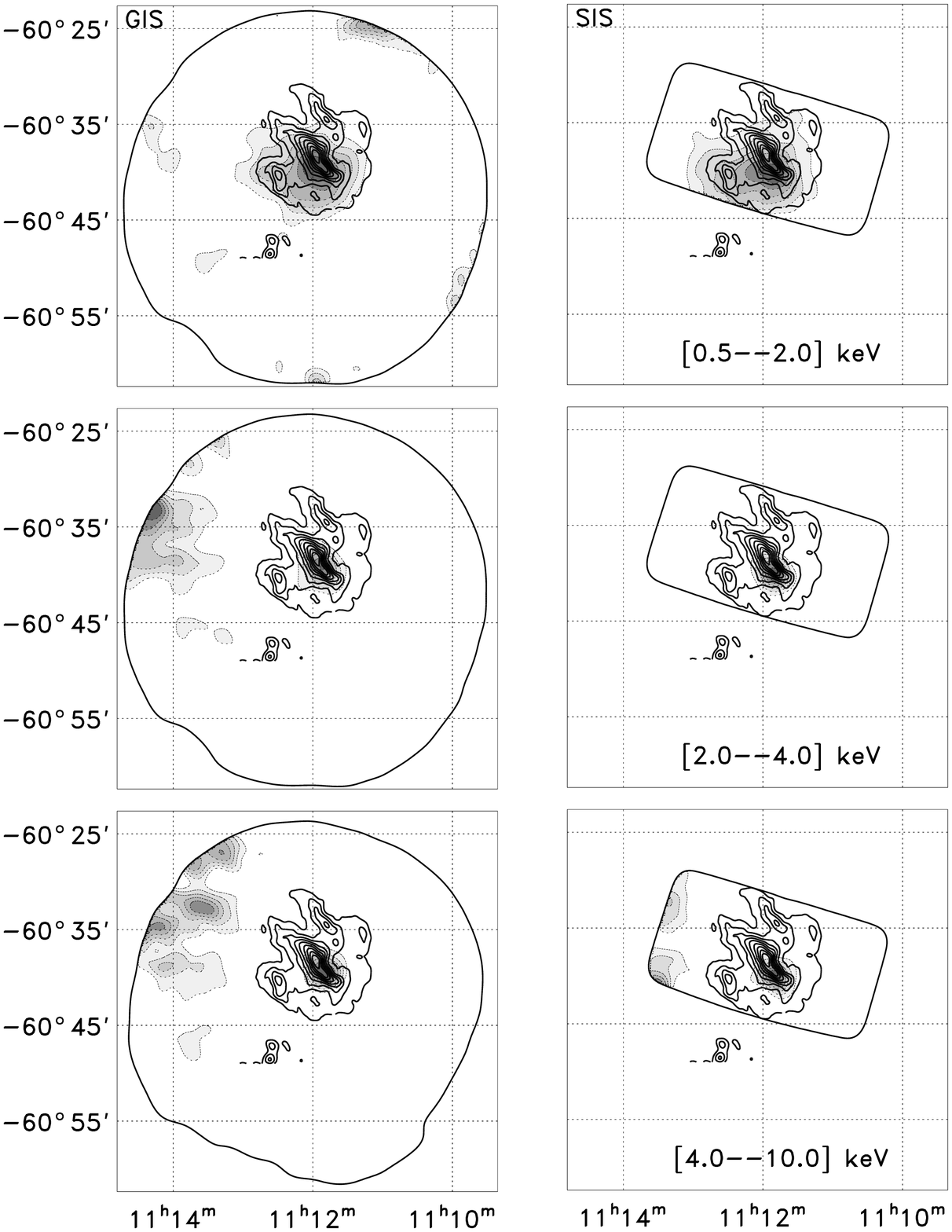,width=6in}}
\end{figure}
 
\newpage
\begin{figure}[t]
\vspace{-1.5in}
\centerline{\psfig{file=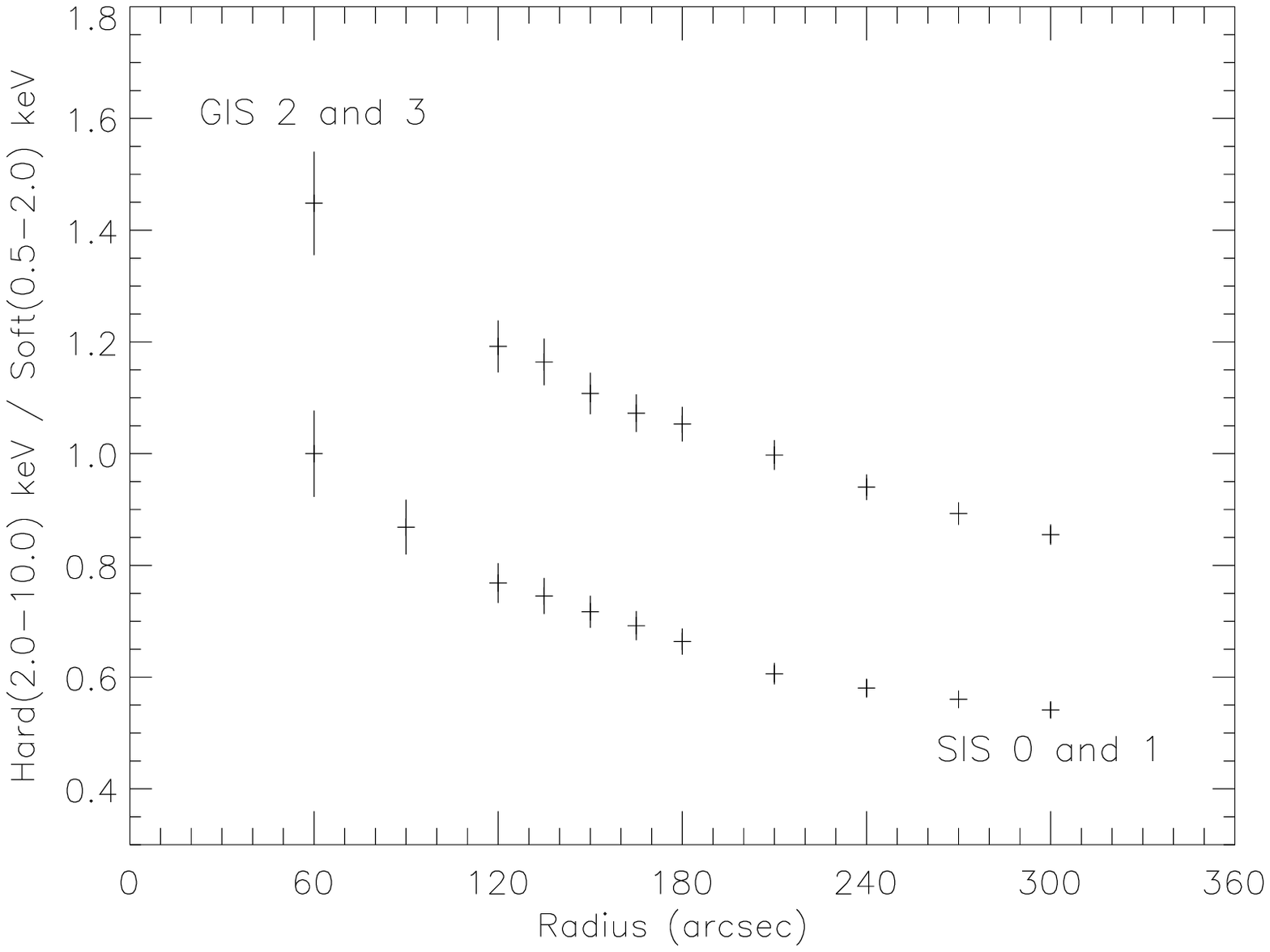,width=6in}}
\end{figure}
\clearpage
 
\newpage
\begin{figure}[t]
\vspace{-1.5in}
\centerline{\psfig{file=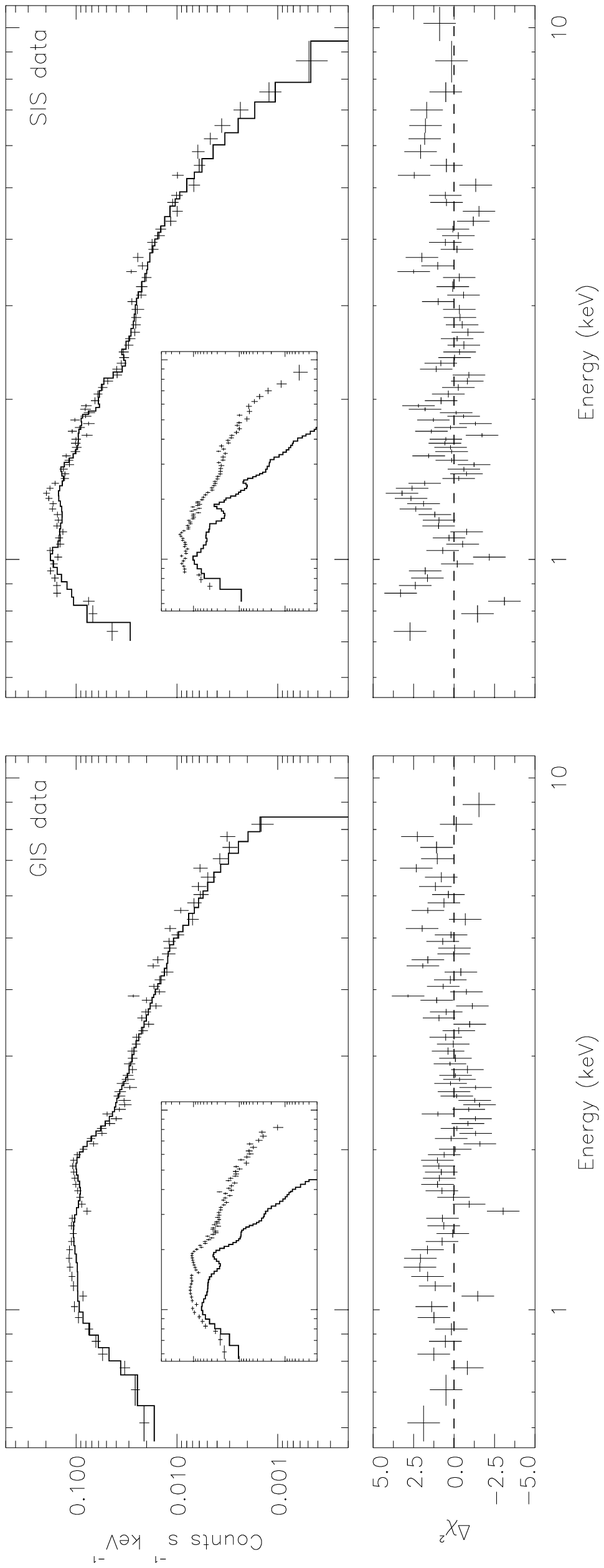,width=6in}}
\end{figure}
 
\newpage
\begin{figure}[t]
\centerline{\psfig{file=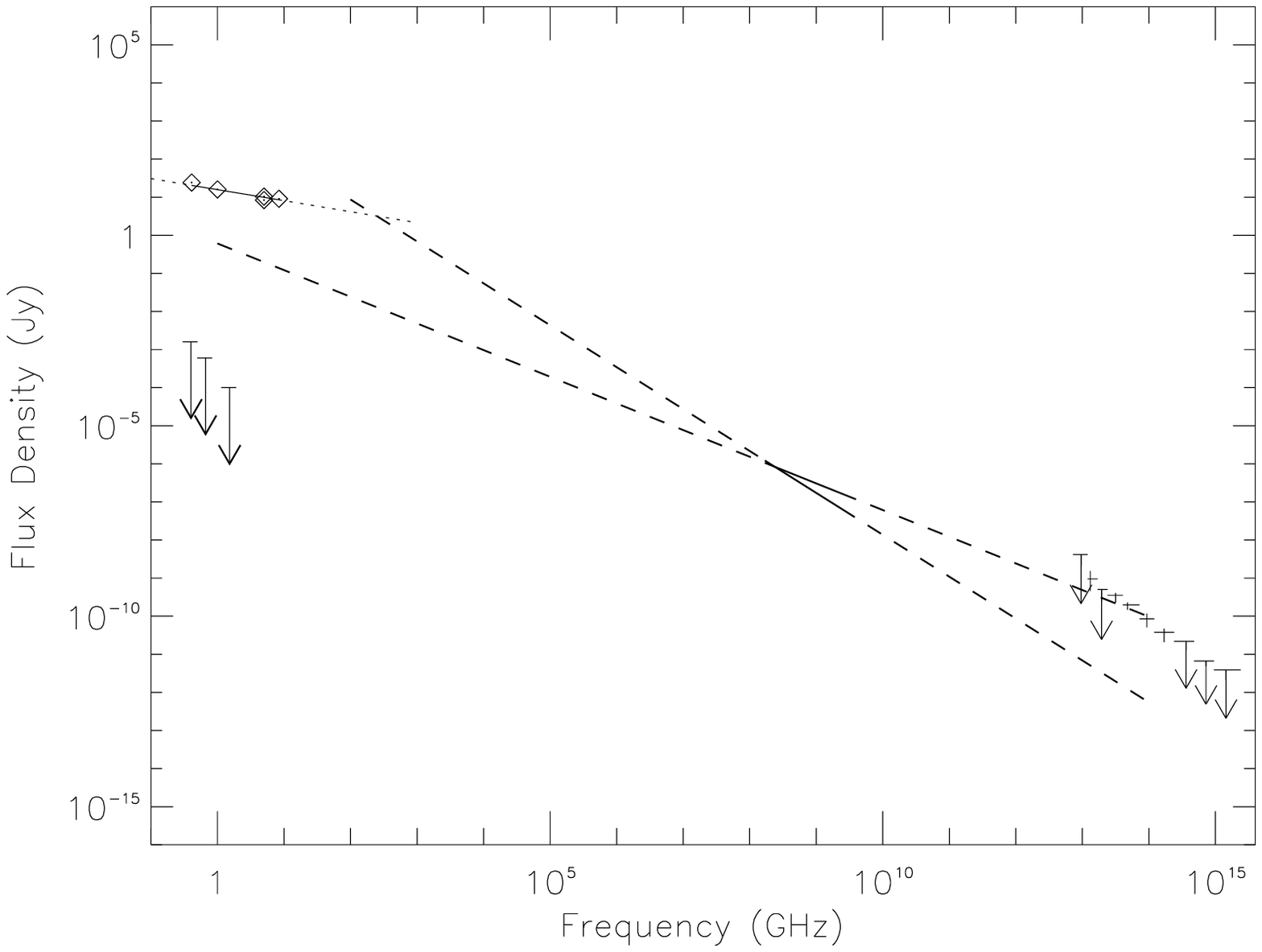,width=6in}}
\end{figure}
\newpage


\begin{thebibliography}{}

\bibitem[]{}
{Anders, E., \& Grevesse, N.} 1989, {Geochimica et Cosmochimica Acta}, { 53}, 197

\bibitem[]{}
{Blanton, E.~L., \& Helfand, D.~J.} 1996, {ApJ}, { 470}, 961

\bibitem[]{}
{Burton, W.~B., \& te Lintel Hekkert, P.} 1986, {A \& A Suppl}, { 427}, 65

\bibitem[]{}
{Clark, D.~H., \& Caswell, J.~L.} 1976, {MNRAS}, { 174}, 267

\bibitem[]{}
{Dickey, J.~M., \& Lockman, F.~J.} 1990, {ARA \& A}, { 28}, 215

\bibitem[]{}
{Ginzburg, V.~L. \& Syrovatskii, S.~I.} 1965, {ARA \& A}, { 3}, 297

\bibitem[]{}
{Green, D.~A.} 1996, {`A Catalogue of Galactic Supernova Remnants (August version)'}

\bibitem[]{}
{Green, D.~A. \& Scheuer, P.~A.~G.} 1992, {MNRAS}, { 258}, 833

\bibitem[]{}
{Harrus, I.~M., Hughes, J.~P., \& Helfand, D.~J.} 1996, {ApJ}, { 464}, L161

\bibitem[]{}
{Kaspi, V.~M. et al.} 1997, {ApJ}, { 485}, 820

\bibitem[]{}
{Kaspi, V.~M., Manchester, R.~N., Johnston, S., Lyne,A.~G., \& D'Amico, N.} 1996, {A~J}, { 111}, 2028

\bibitem[]{}
{Merck, M. et al.} 1996, {A\&A Suppl}, { 120}, 465

\bibitem[]{}
{Milne, D.~K.} 1979, {Aust. J. Phys.}, { 32}, 83

\bibitem[]{}
{Morrison, R. \& McCammon, D.} 1983, {ApJ}, { 270}, 119

\bibitem[]{}
{Raymond, J.~C. \& Smith, B.~W.} 1977, {ApJS}, { 35}, 419

\bibitem[]{}
{Reynolds, S.~P. \& Chevalier, R.~A.} 1984, {ApJ}, { 278}, 630

\bibitem[]{}
{Roger, R. et al.} 1986, {MNRAS}, { 219}, 815

\bibitem[]{}
{Sedov, L.~I.} 1959, {Similarity and dimensional methods in mechanics},
(New York: Academic Press)

\bibitem[]{}
{Seward, F.~D. \& Wang, Z.~R.} 1988, {ApJ}, { 332}, 199

\bibitem[]{}
{Seward, F.~D. et al.} 1972, {ApJ}, { 178}, 131

\bibitem[]{}
{Slane, P.~O., Seward, F.~D., Bandiera, R., Torii, K., \& Tsunemi, H.} 1997, ApJ, { 485}, 221

\bibitem[]{}
{Tamura, K., Kawai, N., Yoshida, A., \& Brinkmann, W.} 1996, {PASJ},{ 48}, L33

\bibitem[]{}
{Taylor, G.~I.} 1950, {Proc Royal Soc London}, { 201}, 159

\bibitem[]{}
{Thompson, D.~J. et al.} 1995, {ApJS}, { 101}, 259


\bibitem[]{}
{Vasisht, G., Aoki, T., Dotani, T., Kulkarni, S.~R., \& Nagase, N.} 1996, {ApJ}, { 456}, L59

\bibitem[]{}
{Weiler, K.~W. \& Panagia, N.} 1980, {A\&A}, { 90}, 269

\bibitem[]{}
{Whiteoak, J.~B.~Z. \& Green, A.~J.} 1996, {A \& A Suppl. Ser.}, { 118}, 329

\bibitem[]{}
{Wilson, A.S.} 1986, {ApJ}, { 302}, 718

\end{thebibliography}
\end{document}